# Scaling rules in the science system: influence of field-specific citation characteristics on the impact of individual researchers

**Rodrigo Costas[a 1], Maria Bordons[a], Thed N. van Leeuwen[b], and Anthony F.J. van Raan[b]**

(a) Instituto de Estudios Documentales en Ciencia y Tecnología (IEDCYT), Spanish Council for Scientific Research (CSIC), Madrid, Spain

(b) Centre for Science and Technology Studies (CWTS), Leiden University, Leiden, The Netherlands

***Abstract***

*The representation of science as a citation-density landscape and the study of scaling rules with the field-specific citation-density as a main topological property was previously analysed at the level of research groups. Here the focus is on the individual researcher. In this new analysis, the size-dependence of several main bibliometric indicators for a large set of individual researchers is explored. Similar results as those previously observed for research groups are here described for individual researchers, reinforcing the idea of prevalent scaling rules in the representation of science as a citation-density landscape. Differences among thematic areas with different citation densities are discussed.*

## Introduction

Science can be considered as a complex system of highly interconnected entities (e.g., researchers, research groups, universities) that produce and transfer knowledge. There is much recent work on the study of networks in science, such as those focused on the study of links among authors, publications and citations (Albert & Barabási, 2002; Dorogovtsev & Mendes, 2002; Leicht, Clarkson, Shedden, & Newman 2007). But there is little work on the study of bibliometric indicators and their statistical properties in the context of science as an interconnected system. Particularly important in large networked systems (Caldarelli, Erzan, & Vespignani, 2004) are the relations between large-scale attributes (in science for instance the citation characteristics of fields) and local patterns (for instance the performance in terms of citation-based impact of individual researchers).

The scaling relationships between number of citations and number of publications have been analysed across countries, research fields and institutes (Katz, 1999, 2000, 2005), as well as across research groups (Van Raan, 2008a). An important finding was that citations increase in a power law relationship with the size of the groups, institutions or nations, and a cumulative advantage effect in the scientific community was observed, that is, a size-dependent Matthew effect (Merton, 1988).

In a series of studies on the statistical properties of bibliometric characteristics of research groups (van Raan, 2006a, b; 2008a) the size-dependent nature of impact was analysed focusing on the differences between top-performance and lower performance groups. The crucial finding was that particularly the lower performance groups have a size-dependent (size of a research group in terms of number of publications) *cumulative* advantage[2] for receiving citations. Two different underlying factors interact: firstly, the fraction of not-cited publications, which decreases for lower-performance groups considerably with size (Van Raan, 2006a)[3]; and secondly, the citation-density of the fields, since research groups working in low field citation-density regions tend to benefit the most from a higher number of publications (Van Raan, 2008a). In the latter publication, the scaling behavior in relation to the size-dependency of the main bibliometric indicators applied to the study of research groups for different levels of field-specific citation densities was analyzed.

[1] Corresponding author, e-mail address: rodrigo.costas@cindoc.csic.es

Following this line of research we continue in this paper our exploration of these interdependencies of the science system as a landscape characterized by field-specific citation densities including a new level of analysis: the individual researcher. We wonder whether the scaling behavior identified at the research group level could also be observed at the individual level. Different questions emerge. Do the scaling rules described for the research groups apply also for individual researchers? Is research performance of researchers influenced by the characteristics of fields, and if so, how?

The structure of this paper is as follows. First, we discuss the data material, the application of the method and the calculation of indicators. Secondly, we present the results of our data analysis for 'external' (i.e., non self-) citations, and finally we discuss the main outcomes of this study in the framework of the landscape model.

## Data, Indicators, Citation-Density Landscape

The data material for this study comes from the analysis of scientific activity of a total of 1,064 researchers working as scientific staff at the Spanish CSIC[4] in 2005. These researchers are grouped according to the thematic orientation of their institutes in three main scientific areas, namely Natural Resources (349 researchers), Biology & Biomedicine (388 researchers) and Materials Science (327 researchers). In total, the analysis involves 1,038 researchers with at least one publication ('CI publications'[5]) in the period under study, and covers about 25,000 publications - published during 1994-2004- and 222,300 citations (excluding self-citations).

Obtaining bibliometric indicators at the individual level is laborious due to the lack of normalization of author names in the publications. A careful analysis of author names and addresses in publications was carried out to identify properly the scientific production of researchers (Costas & Bordons, 2005).

[2] With 'cumulative advantage' we mean that the dependent variable (for instance, number of citations of a group ***C***) scales in a disproportional, non-linear way (power law) with the independent variable (for instance, the 'size' of a research group, in terms of number of publications, ***P***). Thus, larger groups (in terms of ***P***) do not just receive more citations (as can be expected), but they do so increasingly more 'advantageously': groups that are twice as large as other groups receive, for instance 2.4 times more citations. For a detailed discussion we refer to our previous paper (van Raan, 2006b). For a general discussion of cumulative advantage in science we refer to Merton (1988) and Price (1976).
[3] In this context the role of self-citation as impact-reinforcing mechanism is discussed in van Raan (2008b)
[4] Consejo Superior de Investigaciones Científicas (Spanish Council for Scientific Research).
[5] Thomson Scientific, the former Institute for Scientific Information (ISI) in Philadelphia, is the producer and publisher of the Web of Science (WoS) that covers the Science Citation Index (-extended), the Social Science Citation Index and the Arts & Humanities Citation Index. Throughout this paper we use the term 'CI' (Citation Index) for the above set of databases.

The indicators are calculated on the basis of a total time-period analysis. This means that publications are counted for the entire 11-year period (1994-2004) and citations are counted up to and including 2004 (e.g., for publications from 1994, citations are counted from 1994 to 2004; and for publications from 2004, citations are counted only in 2004). We applied the CWTS standard bibliometric indicators. Here only 'external' citations, i.e. citations corrected for self-citations[6], are taken into account. We present the standard bibliometric indicators with a short description in the text box here below. For a detailed discussion we refer to van Raan (2004). For the analysis, only researchers with at least 5 publications where considered.

**Standard Bibliometric Indicators:**

- Number of publications ***P*** in CI-covered journals of a researcher in the specified period;
- Number of citations ***C*** received by ***P*** during the specified period, without self-citations; including self-citations: ***Ci,*** i.e., number of self-citations ***Cs = Ci – C***, relative amount of self-citations ***Cs/Ci***;
- Average number of citations per publication, without self-citations (***CPP***);
- Journal-based worldwide average impact as an international reference level for a researcher (***JCS***, journal citation score, which is our journal impact indicator), without self-citations (on a world-wide scale!); in the case of more than one journal we use the (weighted) average ***JCSm***; for the calculation of ***JCSm*** the same publication and citation counting procedure, time windows, and article types are used as in the case of ***CPP***;
- Field-based[7] worldwide average impact as an international reference level for a researcher (***FCS***, field citation score), without self-citations (on a world-wide scale!); in the case of more than one field (as almost always) we use the (weighted) average ***FCSm***; for the calculation of ***FCSm*** the same publication and citation counting procedure, time windows, and article types are used as in the case of ***CPP***; we refer in this article to the ***FCSm*** indicator as the 'field-specific field citation-density';
- Comparison of the ***CPP*** of a researcher with the world-wide average based on ***JCSm*** as a standard, without self-citations, indicator ***CPP/JCSm***;
- Comparison of the ***CPP*** of a researcher with the world-wide average based on ***FCSm*** as a standard, without self-citations, indicator ***CPP/FCSm***;
- Ratio ***JCSm/FCSm*** is the relative, field-normalized journal impact indicator.

Among these indicators, the internationally standardized (field-normalized) ***CPP/FCSm*** indicator is regarded as the 'crown indicator''. This indicator enables us to observe whether the performance of a unit of analysis is significantly far below (indicator value < 0.5), below (0.5 – 0.8), around (0.8 – 1.2), above (1.2 – 1.5), or far above (>1.5) the international impact standard of the field. Particularly with a ***CPP/FCSm*** value above 1.5, units of analysis can be considered as scientifically strong. A value above 2 indicates a very strong unit and units with values above 3 can generally be considered as excellent and comparable to top-units at the best US universities (van Raan, 2004). A good correlation of ***CPP/FCSm*** and quality judgment of peers has been described elsewhere (Rinia, van Leeuwen, van Vuren, & van Raan 1998, 2001).

In this work, the relationship between variables is studied through correlation analyses comparing the behaviour of *high* and *low* field citation-density researchers in order to detect 'advantages' and 'disadvantages' of the increasing size of production (number of publications) on the impact of research. Due to the high variability of data, a low $R^2$ (determination coefficient) was obtained in some cases and complementary tests were used to support the results. Researchers were grouped into four categories (P1-P4) according to their production and differences by productivity of a number of impact measures were explored for high and low field citation-density researchers with help of ANOVA[8], after normalisation of the variables by the logarithm. For the classification of researchers into four productivity classes the percentiles of the distribution of number of publications within each thematic area are used (see Appendix, Table A1).

[6] A citation is a self-citation if any of the authors of the citing paper is also an author of the cited paper.
[7] We here use the definition of fields based on a classification of scientific journals into *categories* developed by Thomson Scientific/ISI. Although this classification is not perfect, it provides a clear and 'fixed' consistent field definition suitable for automated procedures within our data-system.

## Results and discussion

### *Influence of field-specific citation-density and journal impact*

In Fig. 1 we present the distribution of publications by scientific fields[6] in order to show the thematic composition of each of the three main CSIC areas. Thus, in this study we consider an area as a higher, interdisciplinary aggregate of several fields.

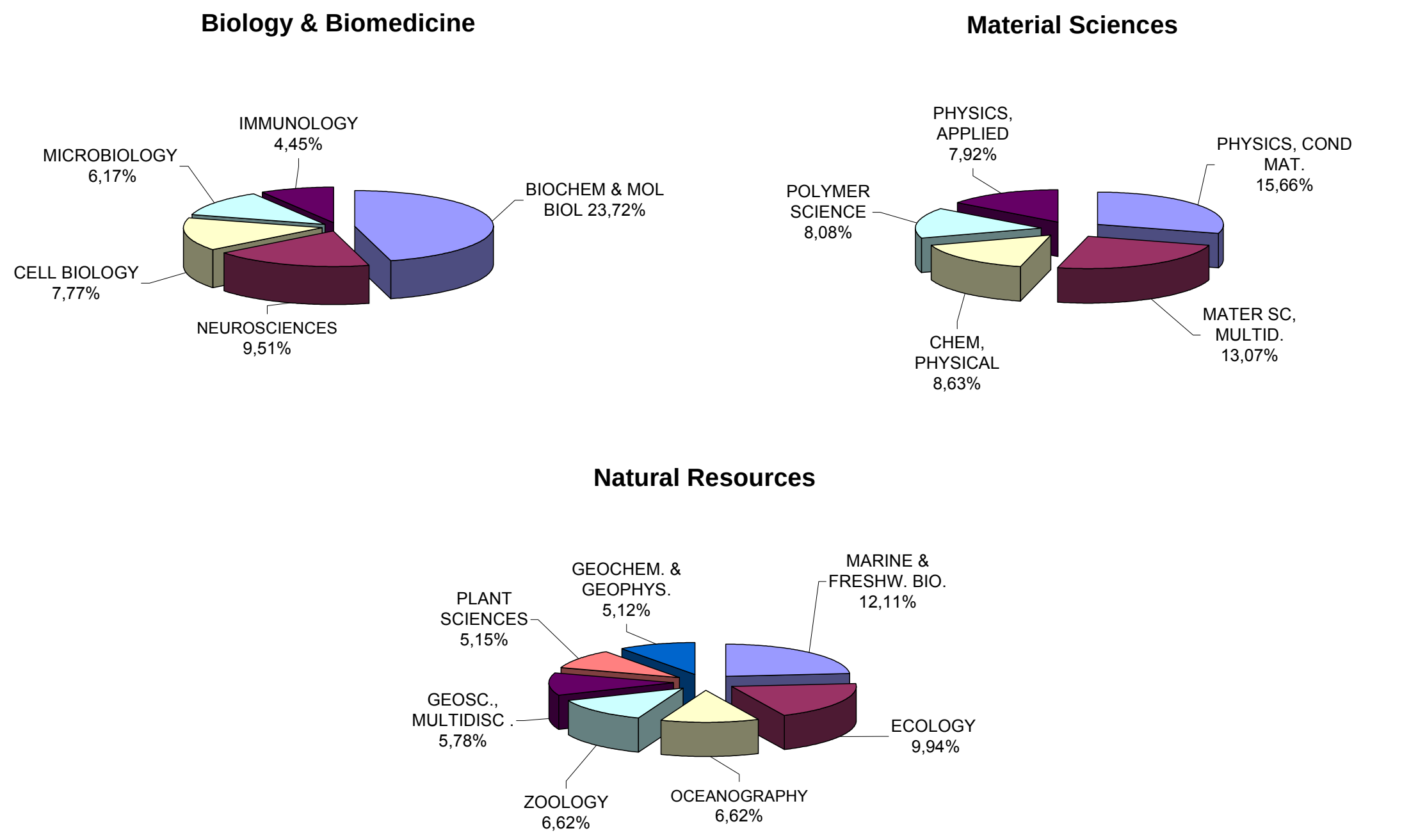

***Figure 1:*** *Distribution of publications within the three areas by field*

In this sense, there are clear scientific orientations in each area. Biology & Biomedicine researchers present a higher percentage of their publications in the fields Biochemistry & Molecular Biology and Neurosciences. Materials Science researchers show an orientation towards Condensed Matter Physics, Materials Science, Physical Chemistry, Polymer Sciences, etc. Natural Resources is the most interdisciplinary area with publications in a wide range of different fields such as Marine and Freshwater Biology, Ecology, Oceanography, Zoology, etc.

---

[8] Analysis of variance (ANOVA) is a statistical procedure in which the observed variance is partitioned into components due to different explanatory variables.

Differences in field citation-density and research performance of researchers in the three areas are also observed (Table 1). The high field citation-density of Biology & Biomedicine (***FCSm***) is remarkable; it is far above the densities in the other two areas. Although Biology & Biomedicine researchers are not the most productive ones in terms of ***P***, they obtain the highest number of ***CPP***, which is, of course, related to the high ***FCSm*** value. In the three areas, researchers tend to publish in journals with an impact above the average in their field (***JCS/FCSm*** >1), although they do not obtain as many citations as their journals (***CPP/JCSm*** <1). The number of citations received is below the average of their field (***CPP/FCSm*** <1) for Materials Science and Natural Resources, and not statistically different from the worldwide average in the case of Biology & Biomedicine. It is interesting to note that 48% of researchers in Biology & Biomedicine show a ***CPP/FCS*** higher than 1, while this percentage is around 37% in the other two areas.

Table 1. Performance of individual researchers by areas

| | ***N*** | ***P*** | ***C*** | ***CPP*** | ***JCSm*** | ***FCSm*** | ***JCS/FCSm*** | ***CPP/JCSm*** | ***CPP/FCSm*** |
|---|---|---|---|---|---|---|---|---|---|
| Biology & Biomedicine | 371 | 24 | 334 | 12.38 | 17.29 | 12.38 | 1.35 | 0.74 | 0.97 |
| Materials Science | 302 | 42 | 149 | 3.47 | 5.41 | 4.65 | 1.22 | 0.73 | 0.85 |
| Natural Resources | 304 | 23 | 116 | 4.97 | 5.40 | 5.64 | 1.00 | 0.84 | 0.86 |

N = number of researchers. All other figures represent the median of the distribution of the indicator values for all researchers per area.

In Fig. 2 the correlation of the number of citations (***C***) with number of publications (***P***) for each of the three areas is presented. Researchers are classified according to *high* and *low* field citation-densities, i.e. the top-25% and bottom-25%, respectively, of the ***FCSm*** distribution. The figure shows that there is a cumulative 'advantage' effect in the three fields, but it is higher for the *low* field citation-density researchers (***C*** increases with ***P*** more for bottom-25% researchers, power law exponent ***α*** between 1.23 and 1.48). These results are consistent with those obtained at the research group level in a previous paper (van Raan, 2008a).

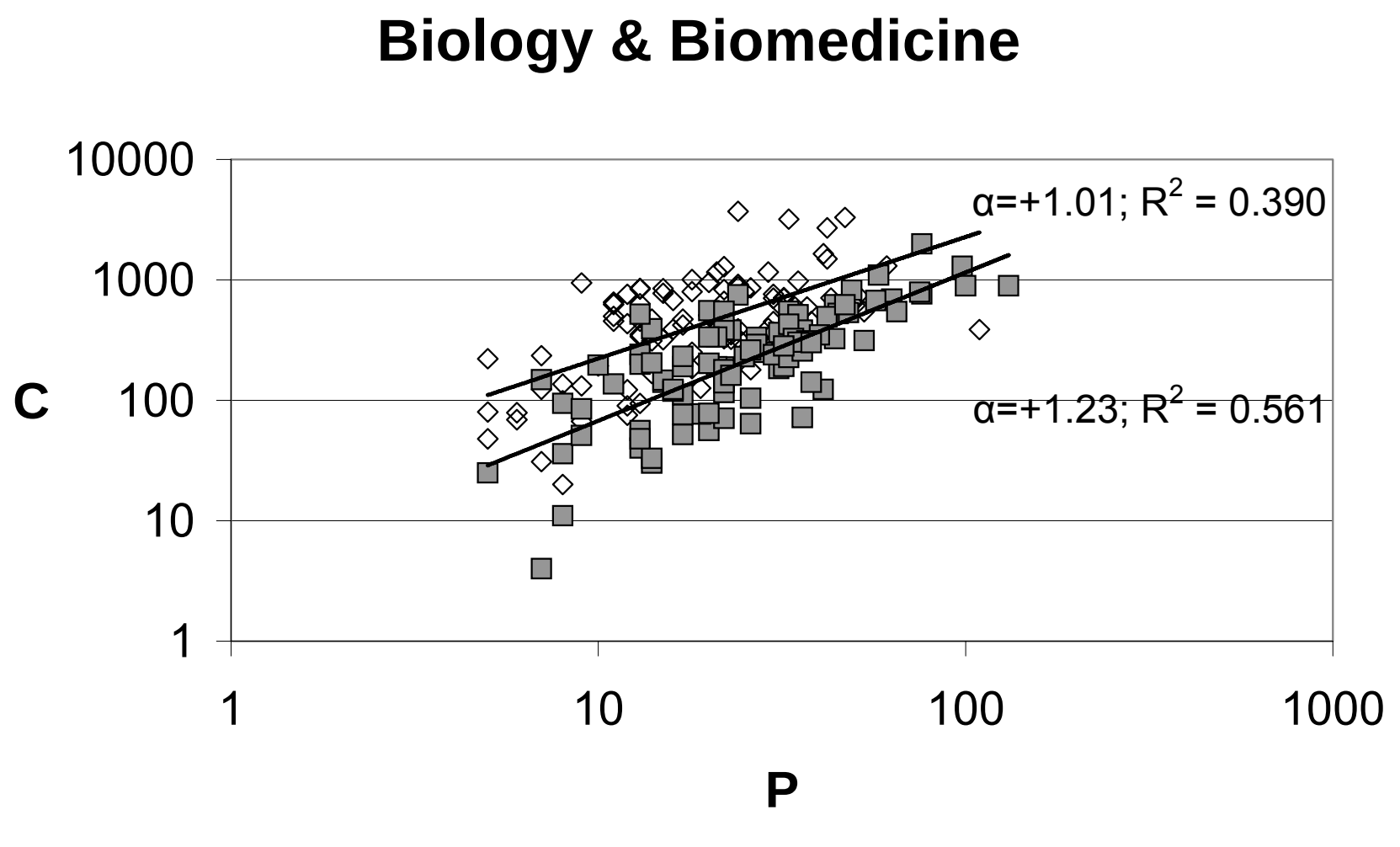

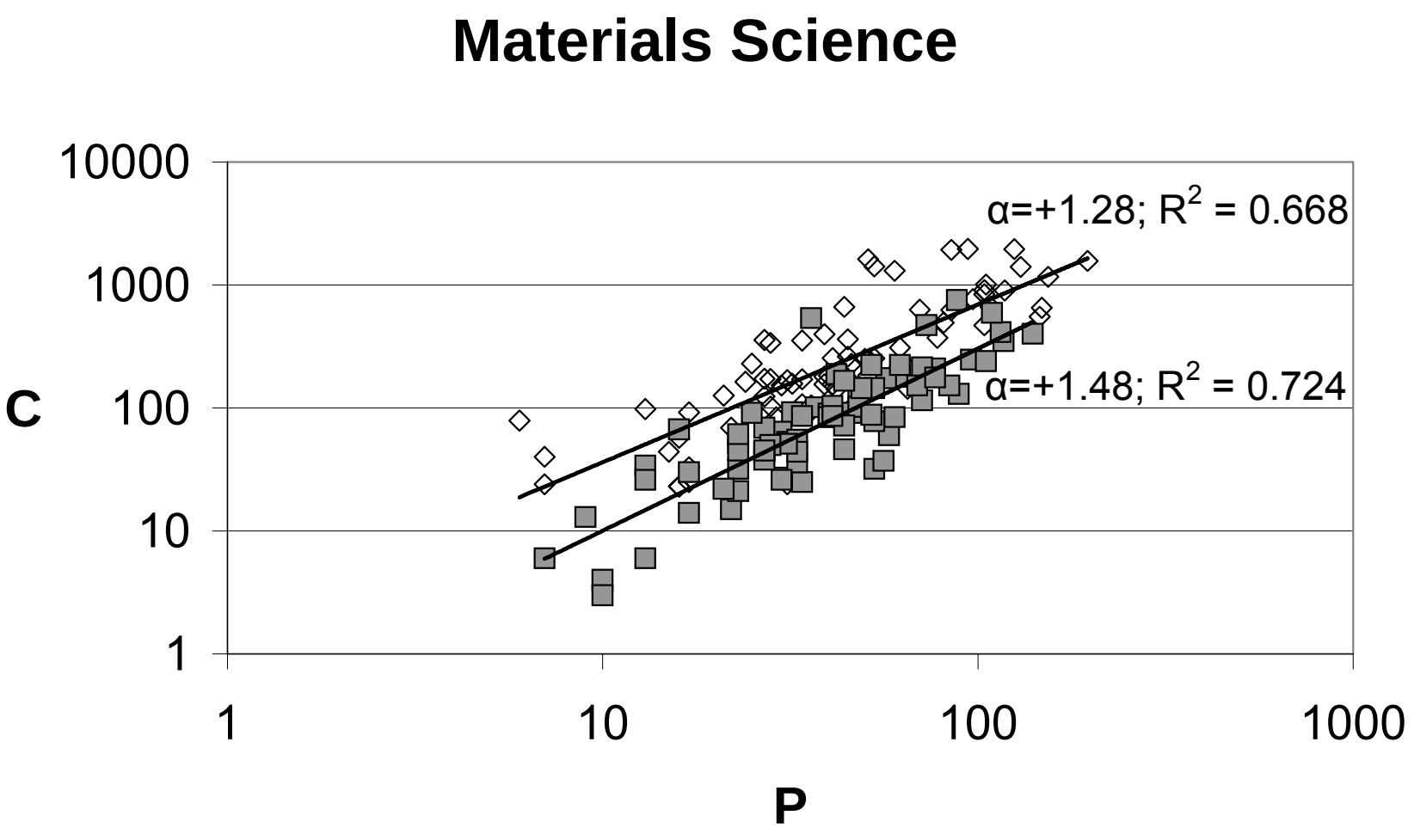

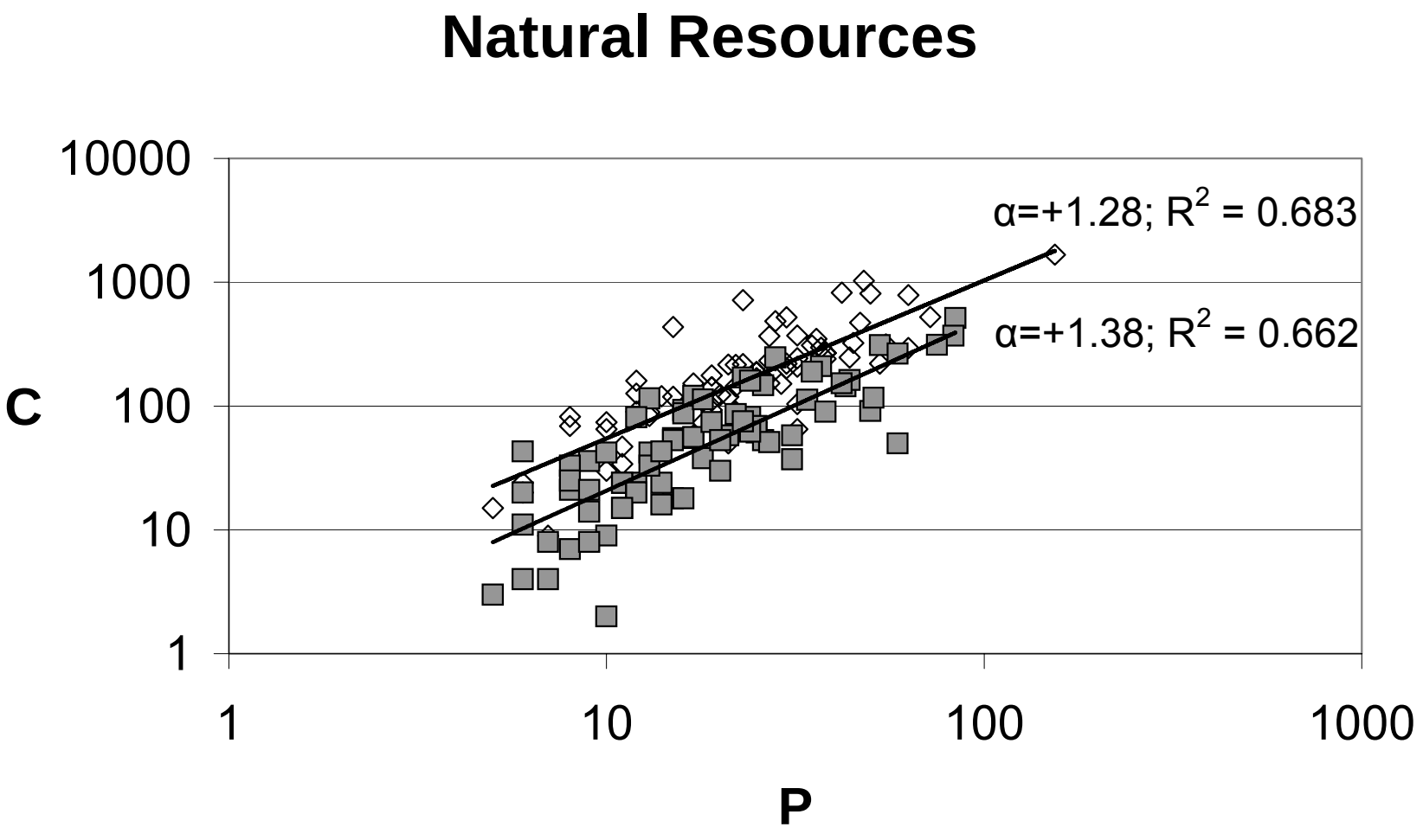

***Figure 2:*** *Correlation of the number of citations (**C**) with the number of publications (**P**), for researchers in the top-25% (diamonds) and in the bottom-25% (squares) of the field citation-density (**FCSm**) distribution.*

Thus, as ***P*** increases, the difference in number of citations between *high* and *low* field citation-density researchers will become smaller. This trend is also observed in Fig. 3 where we show the relationship between ***CPP*** and ***P*** for the high and the low field citation-density researchers. Individual researchers are represented in the figure on the left, while researchers are grouped into four categories according to their production (using the percentiles of ***P***: P1-P4) in the figure on the right. In Materials Science and Natural Resources ***CPP*** tends to increase with ***P***, as observed by the positive power law exponent ***α***, as well as by the fact that high productive researchers (P4) obtain a significant higher ***CPP*** than low productive researchers (P1). Interestingly, the increase of ***CPP*** as a function of ***P*** is higher for *low* field citation-density researchers (higher power law exponent α), which is consistent with previous results for research groups (van Raan, 2008a). However, in that previous study, a slight downward trend of ***CPP*** for increasing values of ***P*** was described for high field citation-density regions.

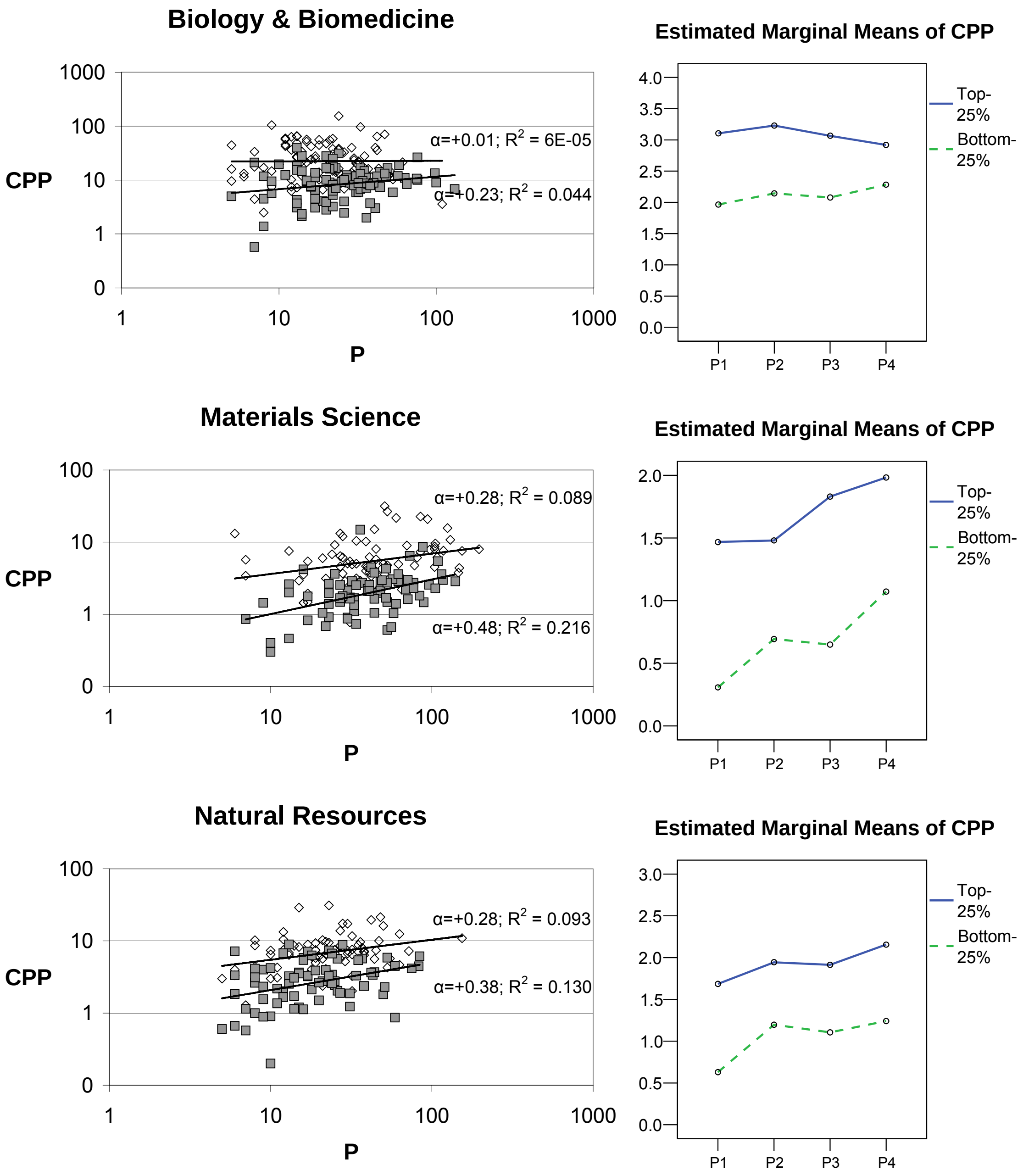

***Figure 3:*** *Correlation of citations-per-publication (**CPP**) with the number of publications (**P**) for researchers in the top-25% (diamonds) and in the bottom-25% (squares) of the field citation-density (**FCSm**) distribution. On the right side we show the results of the percentile-based ANOVA analysis as discussed in the text with the productivity (**P**) percentile class on the horizontal axis, and the natural logarithm (ln) of **CPP** on the vertical axis.*

In Fig. 4 the same data as in Fig. 3 is presented, but now we distinguish within the *high*/*low* field citation-density researchers between top and bottom performance researchers (i.e., the top-50% and the bottom-50% of the ***CPP/FCSm*** distribution, respectively). According to this, four classes can be considered: Top-Top (*High* field citation-density and *top* performance); Top-Bottom (*High* field citation-density and *bottom* performance); Bottom-Top (*Low* field citation-density and *top* performance); and Bottom-Bottom (*Low* field citation-density and *bottom* performance). These data should be analyzed with caution, since we have a low number of researchers in some classes. As a general result, observed in all three scientific areas, the *Bottom-Bottom* researchers benefit more from a larger number of publications, as indicated by their higher power law exponent and the fact that P4 researchers are more productive than P1 researchers ($p<0.05$ in Biology & Biomedicine and Materials Science).

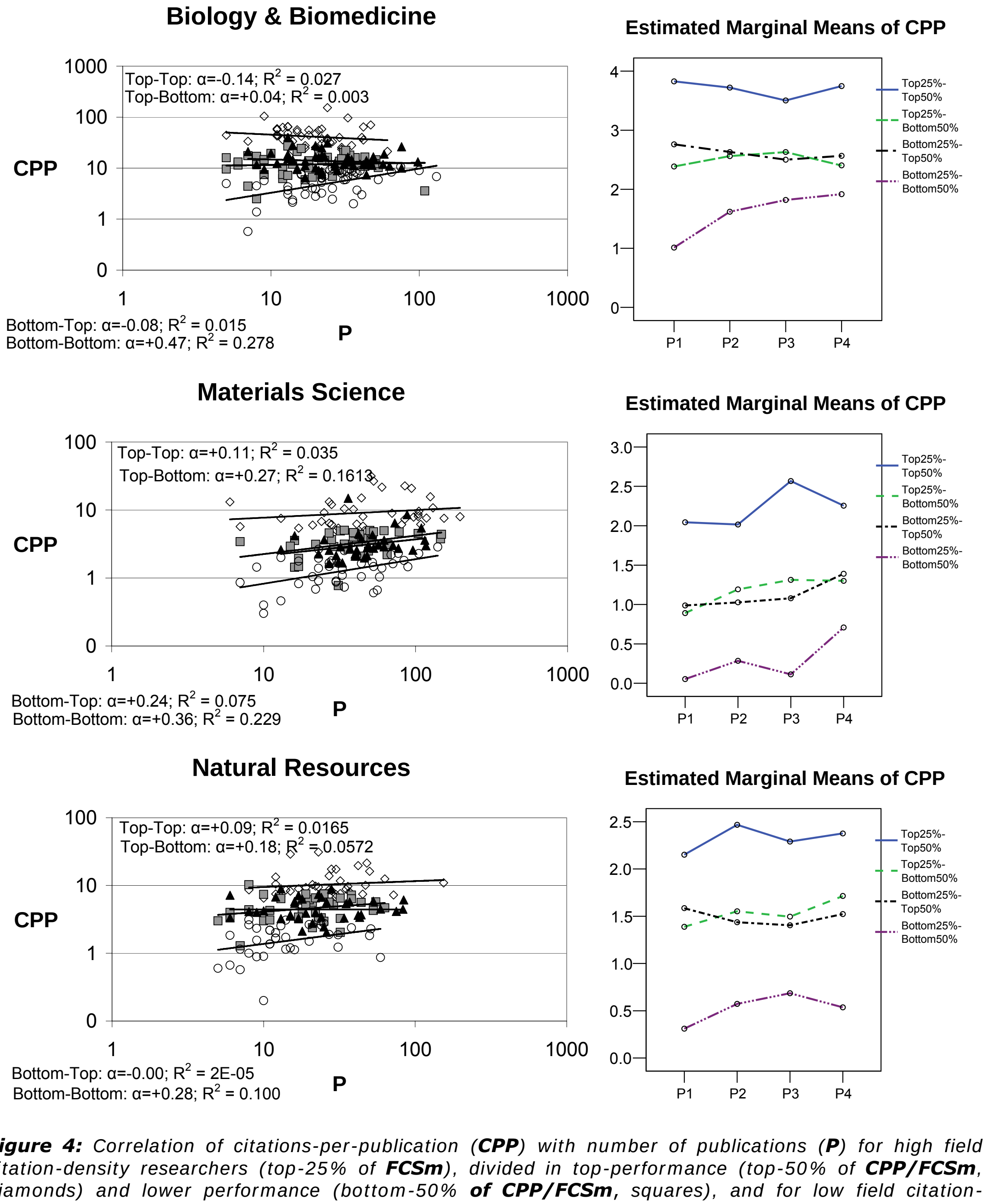

***Figure 4:*** *Correlation of citations-per-publication (**CPP**) with number of publications (**P**) for high field citation-density researchers (top-25% of **FCSm**), divided in top-performance (top-50% of **CPP/FCSm**, diamonds) and lower performance (bottom-50% **of CPP/FCSm**, squares), and for low field citation-density researchers (bottom-25% of **FCSm**), again divided in top-performance (top-50% of **CPP/FCSm**) (triangles) and lower performance (bottom-50% **of CPP/FCSm**, circles). For an explanation of the figures on the right hand side we refer to Fig. 3.*

The behavior of the field citation-density itself as a function of the number of publications for both the high as well as the low field citation-density regions has been investigated. The results are shown in Fig. 5 and similar properties to those previously described for research groups are observed (van Raan, 2008a).

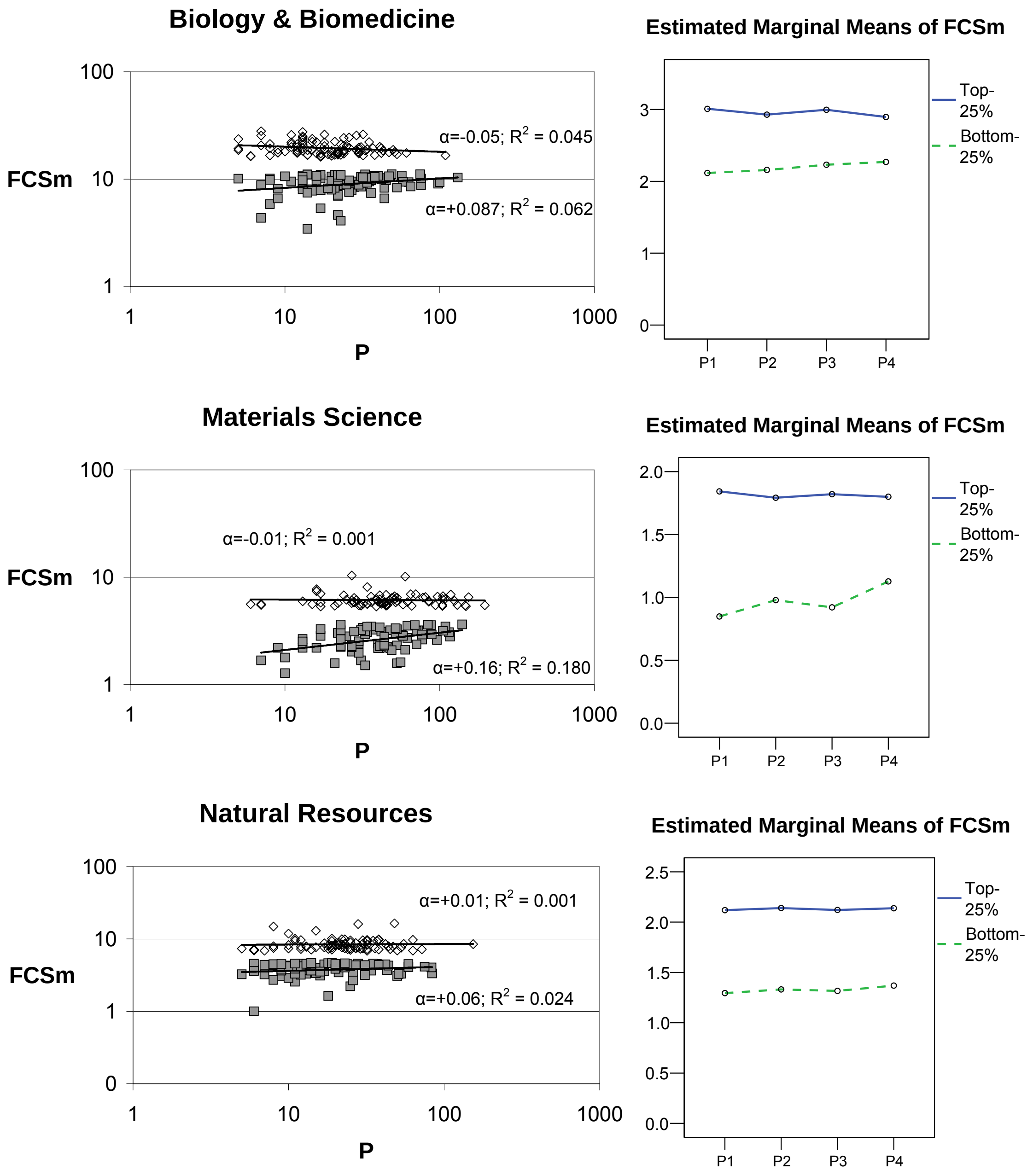

***Figure 5****:* Correlation of field citation-density (***FCSm***) with size (***P***) for researchers in fields with a high (diamonds) and a low (squares) field citation-density (***FCSm***). *On the right side.we show the results of the percentile-based ANOVA analysis as discussed in the text with the productivity (**P**) percentile class on the horizontal axis, and the natural logarithm (ln) of **FCSm** on the vertical axis.*

For the *high* field citation-density researchers the ***FCSm*** tends to decrease very slightly or remains stable with increasing ***P***. Statistically significant results are only found in Biology & Biomedicine, in which the most productive researchers show a lower ***FCSm*** than the least ones ($p<0.01$). This means that for researchers operating in *high* field citation-density regions in this area a larger number of publications mostly implies extension toward regions with a somewhat lower field citation-density.

However, for the *low* field citation-density researchers, we notice a slight upward trend in ***FCSm*** as ***P*** increases. This finding is supported by the observation of a higher ***FCSm*** for the most productive researchers as compared to the least ones (significant differences in Biology & Biomedicine and Materials Science). Thus, for researchers operating in the *low* field citation-density regions, a larger number of publications appears to go together with an 'expansion' into regions with higher field citation-density.

The difference between top and lower performance researchers is shown in Fig.6. We clearly observe, particularly in the right hand side figures, that there is no significant difference between top and low performance as was also found in the case of groups (van Raan 2008a).

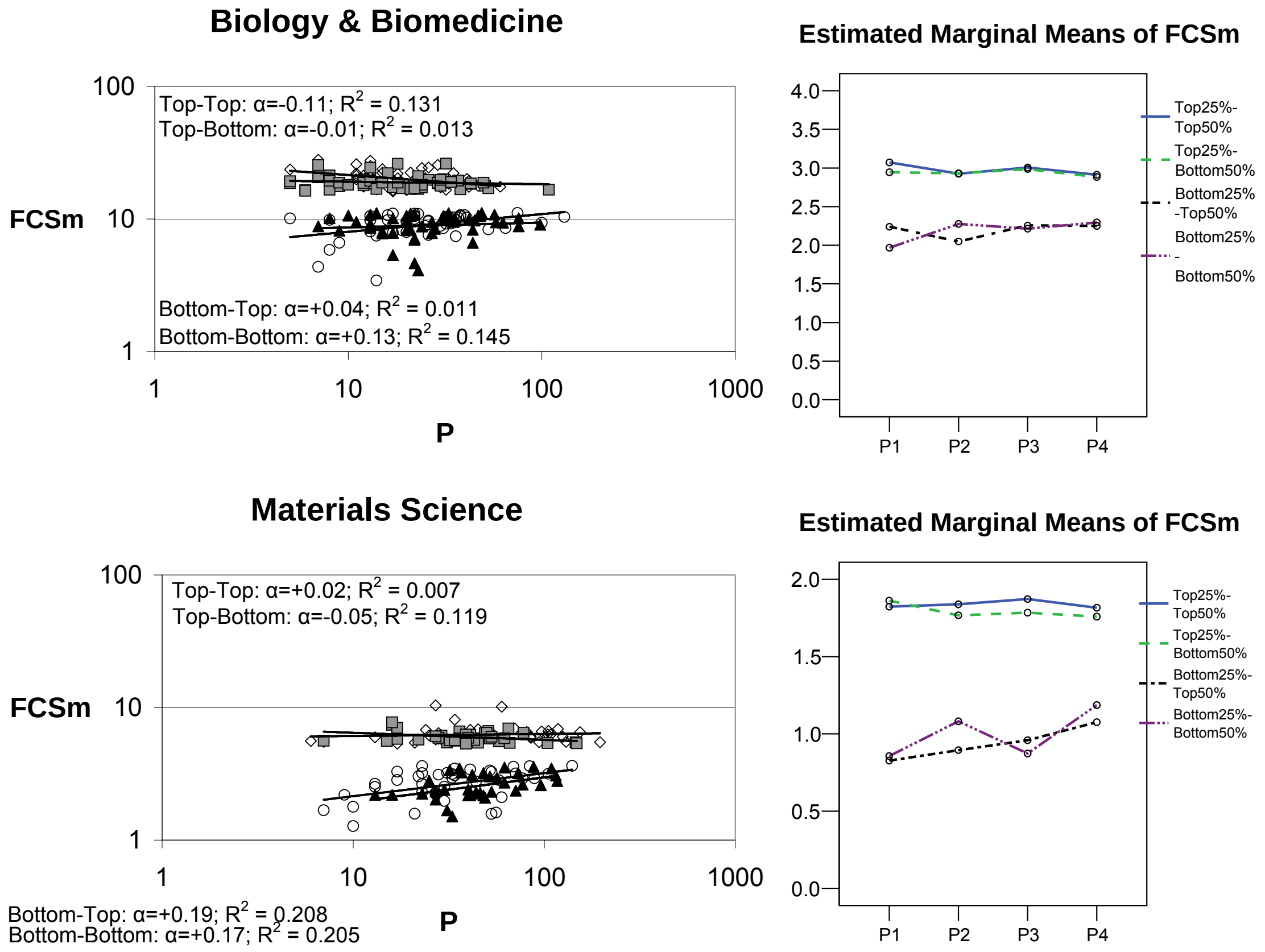

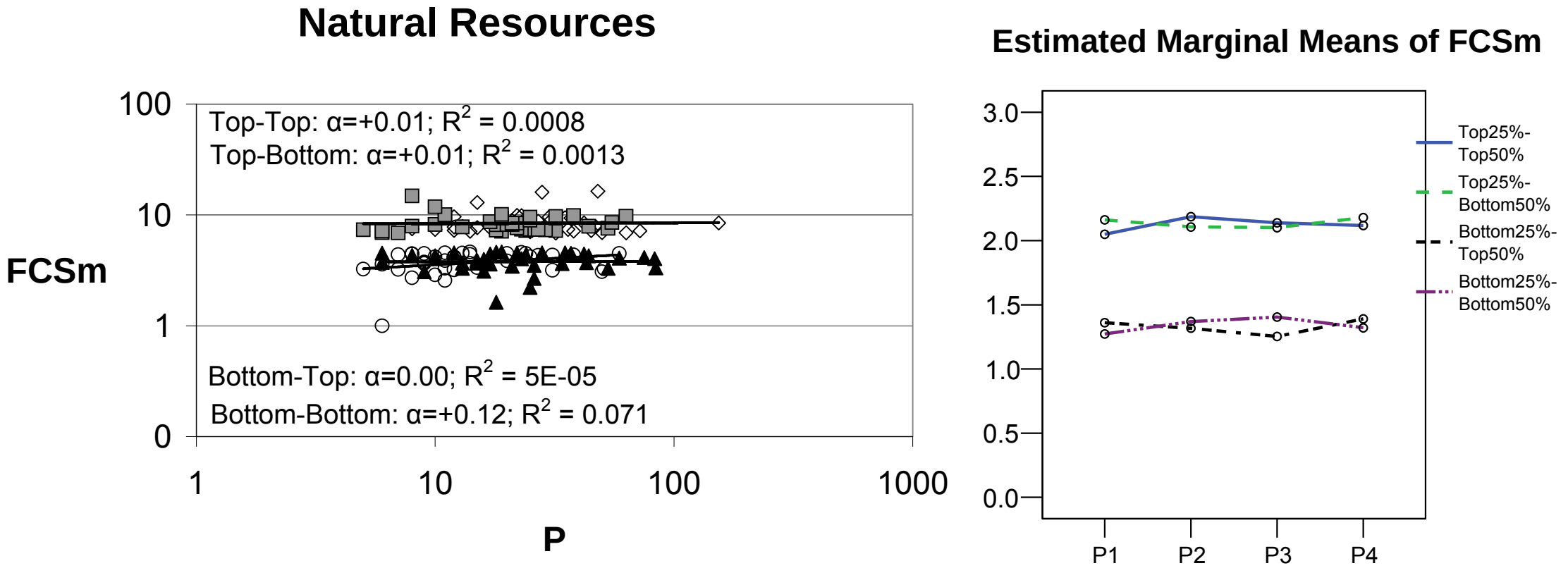

***Figure 6:*** *Correlation of field citation-density (**FCSm**) with number of publications (**P**) for high field citation-density researchers (top-25% of **FCSm**), divided in top-performance (top-50% of **CPP/FCSm**, diamonds) and lower performance (bottom-50% **of CPP/FCSm**, squares), and for low field citation-density researchers (bottom-25% of **FCSm**), again divided in top-performance (top-50% of **CPP/FCSm**) (triangles) and lower performance (bottom-50% **of CPP/FCSm**, circles). For an explanation of the figures on the right hand side we refer to Fig. 5.*

How does the average journal citation-impact of a researcher relate to the field citation-density? The answer to the question is given by Fig. 7. We find that for the *low* field citation-density researchers, a larger production implies a higher average ***JCSm*** value. More specifically, the differences in ***JCSm*** of researchers in relation to productivity levels (P1-P4) are significant in the case of Materials Science and Natural Resources ($p<0.05$, figures on the right hand side of Fig. 7). Concerning high field citation-density regions a larger number of publications do not significantly change the average journal citation impact, and even the trend is slightly negative in some cases. This means that 'expanding in size' may take place with the same field citation-density region, publishing in journals with the same or lower impact.

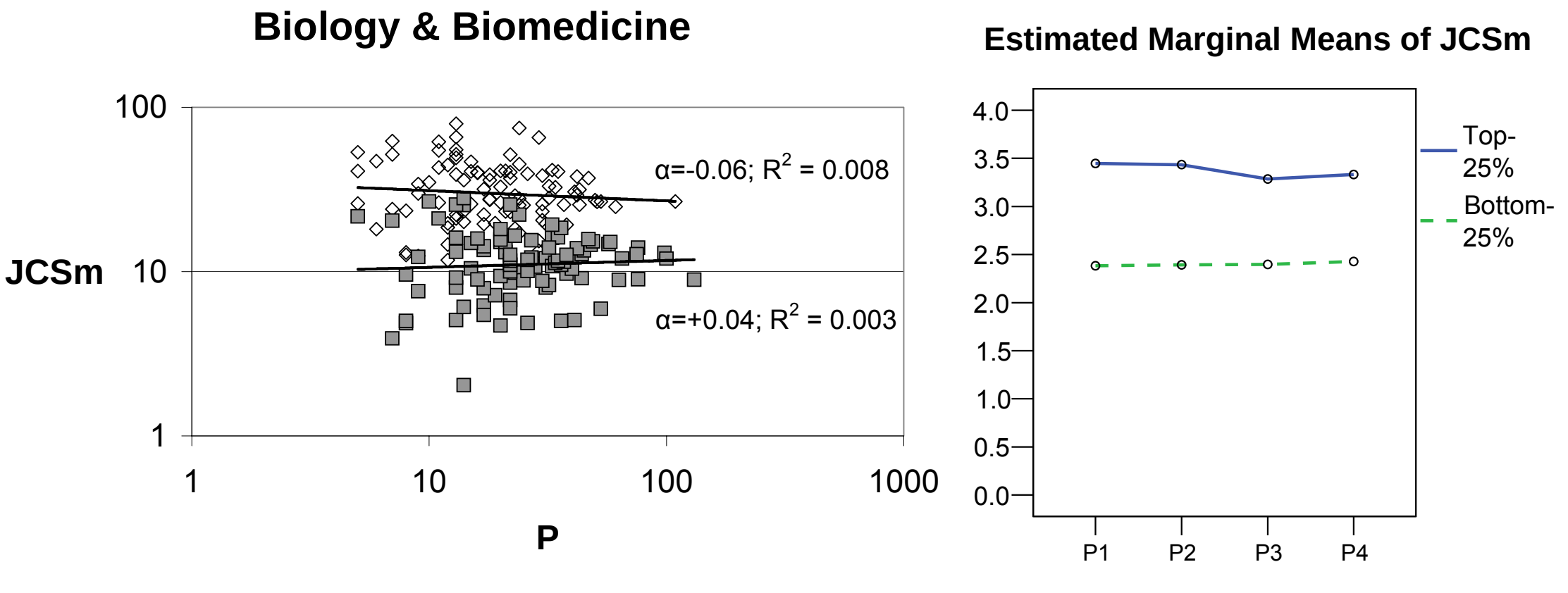

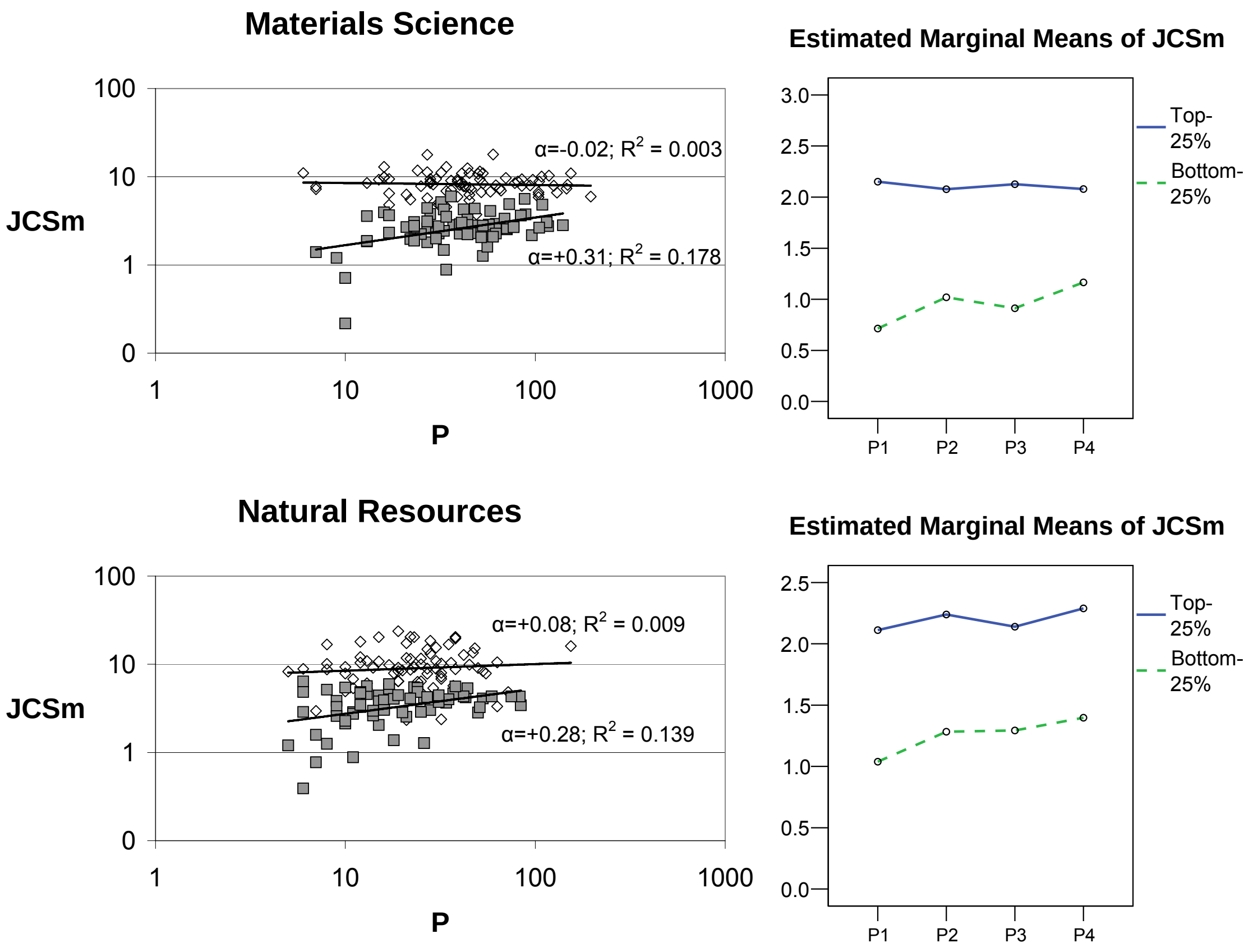

***Figure 7**: Correlation of journal impact (**JCSm**) with size (**P**) for researchers in fields with a high (diamonds) and a low (squares) field citation-density (**FCSm**). On the right side we show the results of the percentile-based ANOVA analysis as discussed in the text with the productivity (**P**) percentile class on the horizontal axis, and the natural logarithm (ln) of **JCSm** on the vertical axis.*

Fig. 8 shows that the lower performers in *low* field citation-density fields (Bottom-Bottom) are the ones who benefit the most: for them a larger number of publications implies higher ***JCSm*** scores in the three areas analysed. In fact, significant differences in ***JCSm*** by productivity classes are found for bottom-bottom researchers in Biology & Biomedicine and Materials Science ($p<0.05$ and $p<0.01$, respectively).

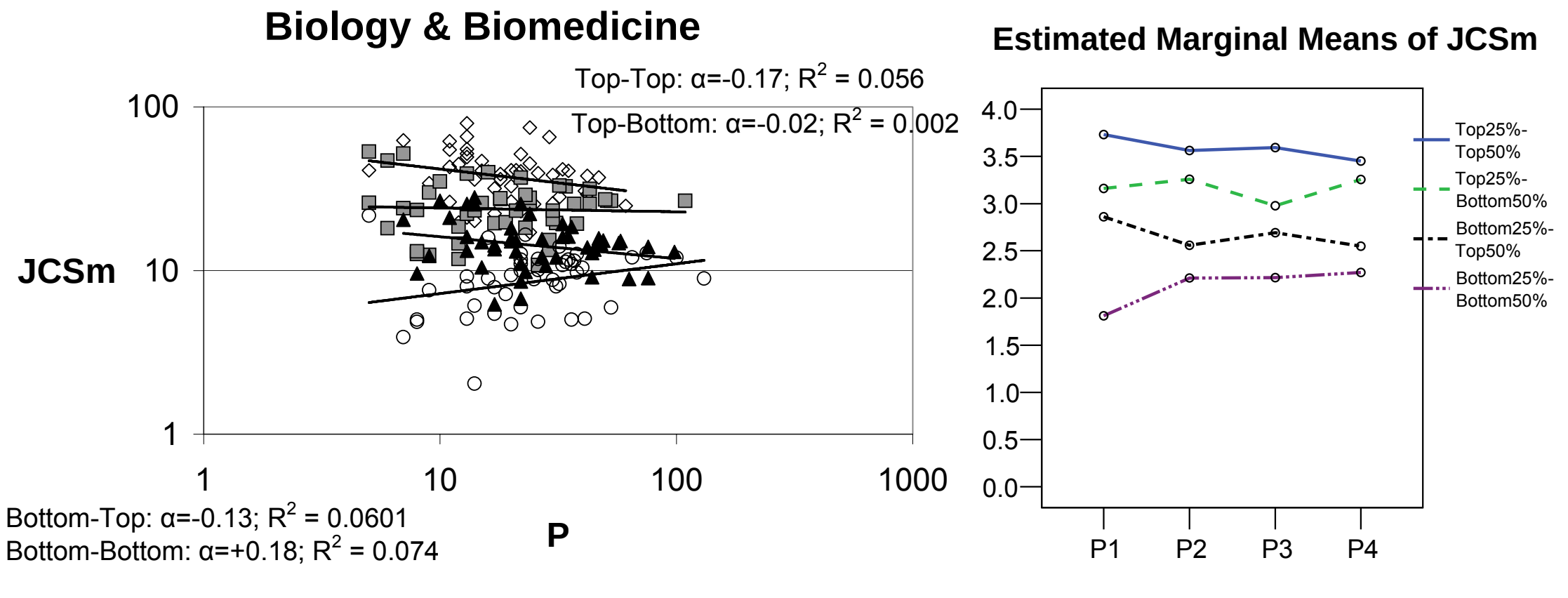

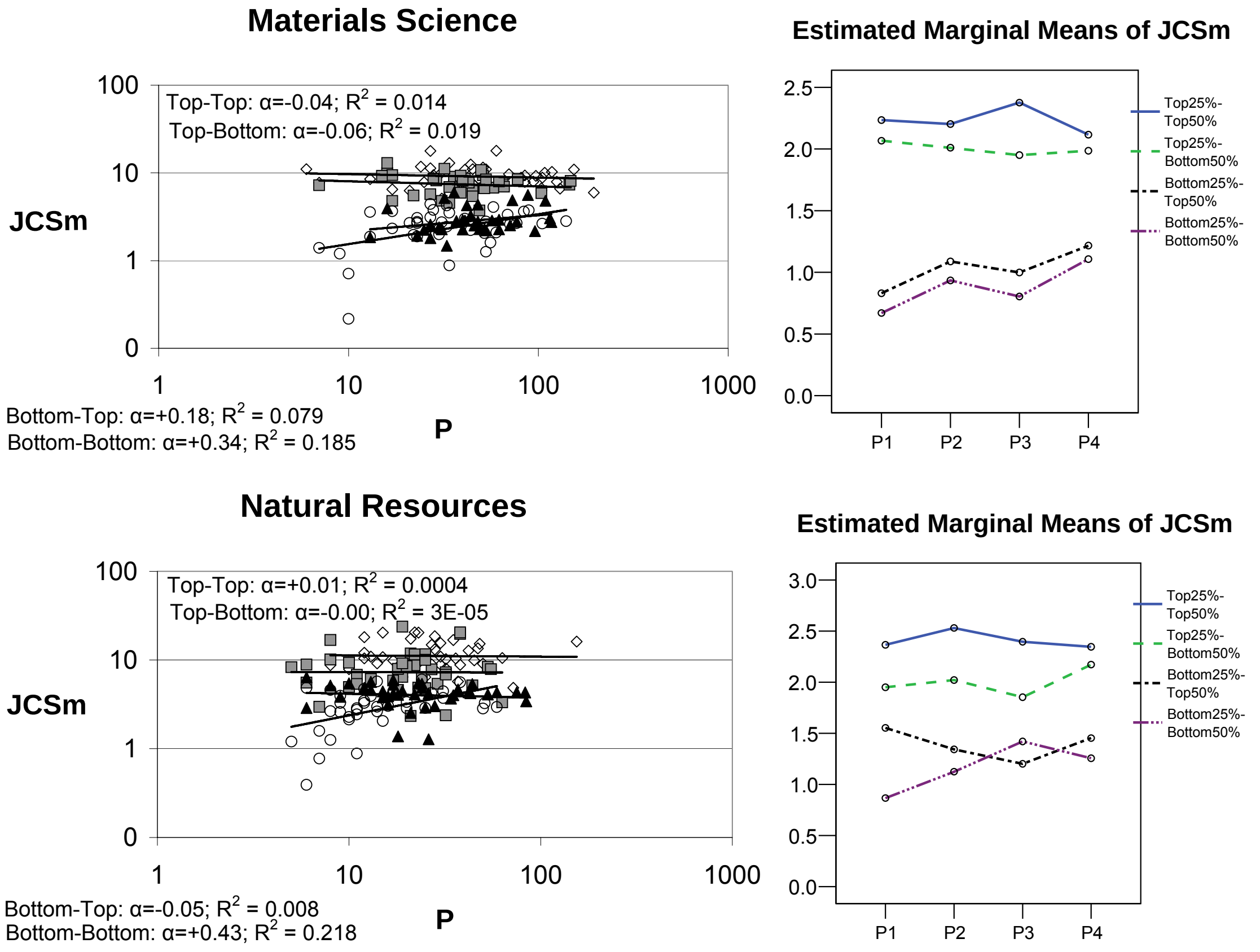

***Figure 8:*** *Correlation of field citation-density (**JCSm**) with number of publications (**P**) for high field citation-density researchers (top-25% of **FCSm**), divided in top-performance (top-50% of **CPP/FCSm**, diamonds) and lower performance (bottom-50% **of CPP/FCSm**, squares), and for low field citation-density researchers (bottom-25% of **FCSm**), again divided in top-performance (top-50% of **CPP/FCSm**) (triangles) and lower performance (bottom-50% **of CPP/FCSm**, circles). For an explanation of the figures on the right hand side we refer to Fig. 7.*

Thus, for researchers operating in *low* citation-density regions a larger number of publications can be seen as an 'expansion' into regions with *higher* field citation-density as we saw earlier and at the same time as an expansion towards journals with a higher average impact.

The interrelation between field citation-density and journal impact, and its influence on the total number of citations of a researcher, also needs to be studied. We take the results presented in Fig. 2 and make a breakdown for both the *high* (top-25% of the ***FCSm*** distribution) as well as the *low* (bottom-25% of ***FCSm***) into the higher (top-50% of the ***JCSm*** distribution) and the lower (bottom-50% of ***JCSm***) journal impact, see Fig. 9.

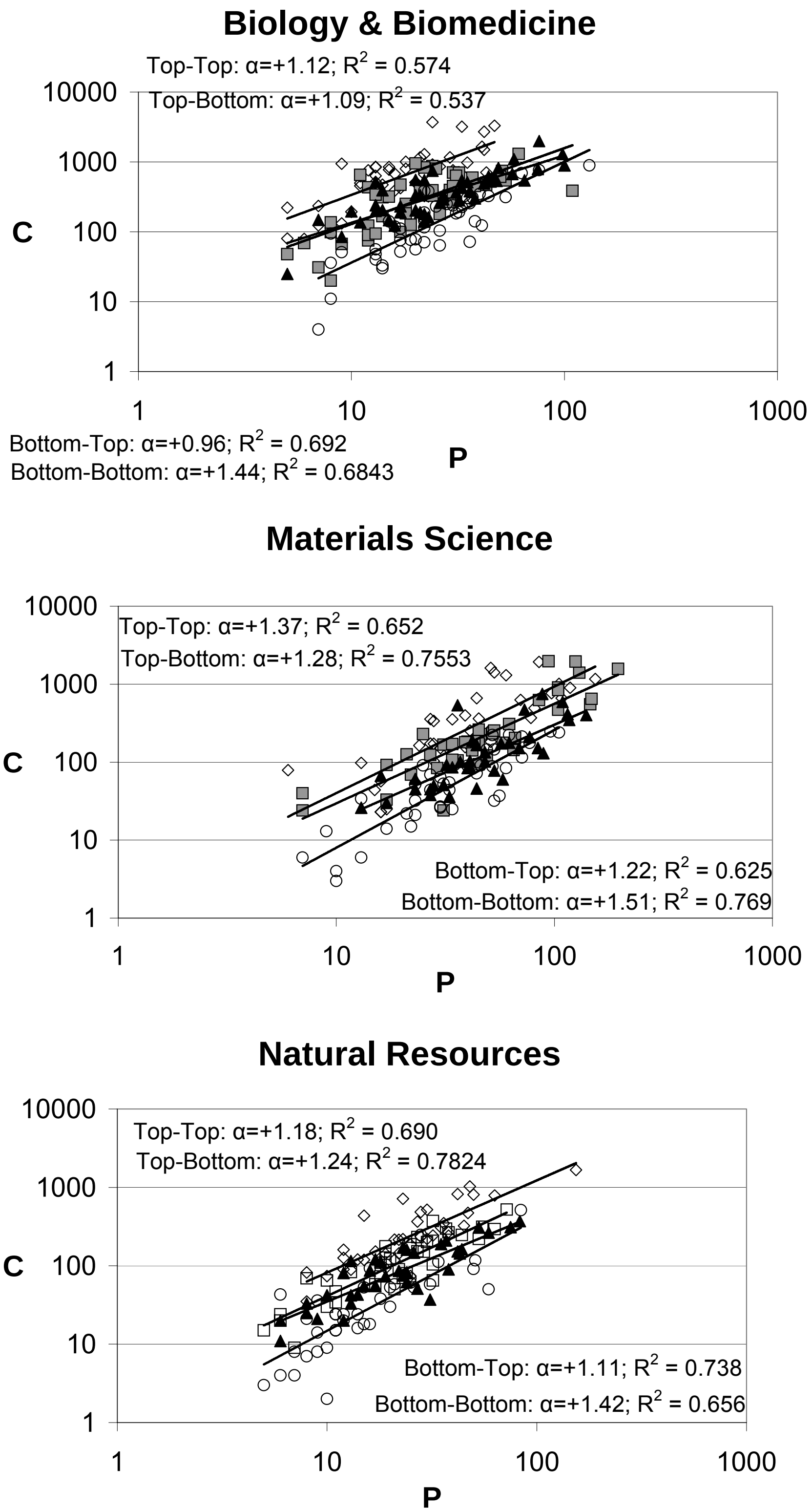

***Figure 9:*** *Correlation the number of citations (**C**) with the number of publications (**P**) for high field citation-density researchers (top-25% of **FCSm**), divided in high journal impact (top-50% of **JCSm**, diamonds) and low journal impact (bottom-50% **JCSm**, squares), and for low field citation-density researchers (bottom-25% of **FCSm**), again divided in high journal impact (top-50% of **JCSm**) (triangles) and low journal impact (bottom-50% **JCSm**, circles).*

Clearly we observe that researchers in low field citation-density fields who publish in *low* impact journals (Bottom-Bottom) benefit the most from the increase in number of publications in the three areas (power law exponent $\alpha$ is between +1.42 and +1.51).

## Analysing the observations in the framework of the science landscape

In a previous paper (van Raan, 2008a) the size-dependence nature of several bibliometric indicators was analyzed for a large set of research groups. One of the main observations was that groups in *low* density citation fields have a size-dependent cumulative advantage for receiving citations, and they benefit the most from an increase in the number of publications. Does this behaviour also apply to a lower aggregation level, namely individual researchers? Moreover, since differences between thematic areas can be expected, the performance of individual researchers in three different areas is studied in this paper.

In the comparison of present results with previous ones, several facts need to be considered. Firstly, it must be kept in mind that different thematic areas are analysed: Chemistry in the previous study versus Biology & Biomedicine, Materials Science, and Natural Resources in this paper. Thus, area-specific features in research performance should be taken into account in the understanding of the results for the different areas. Secondly, differences in the organization of research in the Netherlands (previous paper) and Spain (this paper) also could influence the results. Concerning methodological issues, the fact that the same source data and indicators are used in both studies makes comparisons possible.

We are especially interested in the aggregation level-specific differences between groups (previous study) and individual researchers (this study). Some features of the study at the individual level need to be taken into account, such as the higher variability of data and the fact that different researchers from the same group (sharing similar bibliometric features) are playing in the analysis, while at the aggregation level of groups only one value for the bibliometric indicators appears in the analysis. Comparison of the statistical properties of the bibliometric performance indicator values of researchers and groups within the same population should be addressed in future work to answer these questions.

In spite of these limitations, it is fascinating to see that some of the patterns previously observed at the group level do also emerge here at the level of the individual researcher. We show in this paper that the total number of citations received by researchers increases in a cumulatively advantageous way as a function of number of publications with a higher benefit for researchers publishing in fields of *low* field citation-density. The number of citations per publication (***CPP***) also tends to increase with the number of publications (***P***), and this increment is higher for *low* field citation-density researchers, but there is no cumulative advantage (power law exponent <1).

As the production of researchers increases, a trend to publish in higher impact journals (***JCSm***) is more significant for *low* field citation-density researchers as compared to high field citation-density researchers. The trend to publish in fields with a higher density citation (***FCSm***) as a function of the increasing number of publications is weak, but it is again higher for researchers in *low* density citations fields. Thus, for researchers in the latter fields, a larger number of publications imply a higher probability of expansion into higher citation densities. These trends are consistent with results shown at the research group level (van Raan, 2008a), although specific differences in the power law exponent depending on the area can be observed.

According to our results, Materials Science researchers working in *low* field citation-density fields have a stronger advantage with size, since they show the highest power law exponent in the correlation of the different indicators (***C***, ***CPP***, ***FCSm***, ***JCSm***) with ***P*** (see Appendix, Table A2). The other side of the spectrum is Biology & Biomedicine, which shows the lowest exponents, although also higher for *low* as compared to *high* field citation-density regions.

For Materials Science researchers active in *low* density regions, citations and ***CPP*** tend to increase with the number of publications, and they tend to publish in higher impact factor journals and even expand to higher field citation-density regions with increasing production. For Natural Resources researchers working in *low* density regions, citations and ***CPP*** also tends to increase with ***P***, and better journals within the same field are used as scientific production increases. However, in Biology & Biomedicine only researchers working in *low* field citation-density regions and publishing in below- average impact journals show an increase in ***CPP*** for lager ***P***. The fact that Biology & Biomedicine shows the highest density of citations and **CPP*/FCSm*** score (Table 1) might explain its peculiarities. It is a very competitive area and researchers are oriented towards high impact journals within their subfields.

Summarizing, we show that researchers in *low* field citation-density fields benefit the most from increasing number of publications, as previously observed at the level of research groups. However, the difference in advantage between *low* and *high* field citation-density researchers is smaller here than in the study of groups. In fact, a negative effect of the number of publications on ***CPP*** or ***JCS*** was observed at the level of groups in high field citation-density regions (Van Raan, 2008a), so ‘expanding in size’ could be counterproductive in ***CPP*** scores for these groups.

With regard to our results at the individual level, almost no size-dependence effect on ***JCS*** (all three areas) and a positive size effect on ***CPP*** (in Materials Science and Natural Resources) were observed. Since CSIC researchers have, on average, impact scores below the international level (CPP/FCSm<1) (Table 1), we hypothesize that these researchers still have 'room' for improvement (especially in Natural Resources and Materials Science). The fact that Biology & Biomedicine researchers are working in higher field citation-density regions (***FCSm***=12.38 as compared to values below 5 for the other two areas) but manage to publish in high impact journals within the field (***JCS/FCSm***=1.35) suggests that for them improvement is more difficult. In fact, the hypothesis that researchers with ***CPP/FCSm***>1 are less likely to benefit in their ***CPP*** when increasing their number of publications is supported by data in the Appendix, Table A3.

Our results show the existence of a size-dependent cumulative advantage for receiving citations, which has been previously described at the country, institution and group level, and is here also observed at the individual level. Researchers in low field citation density regions and those whose impact is below world class tend to benefit the most from an increase in number of publications. Inter-area differences can be explained by different factors such as the field-citation density and the distribution of high/low performance of researchers as compared to the world average.

### *Acknowledgements*
This work has been possible thanks to an I3P fellowship from the CSIC and a research stay of Rodrigo Costas at the CWTS (Leiden University). Authors also want to thank Laura Barrios (CSIC) for her advice on statistical issues.

### *References*

Albert, R. and A.-L. Barabási (2002). Statistical mechanics of complex networks. *Rev. Mod. Phys.*, 74, 47-97.

Costas, R.; Bordons, M. (2005). Bibliometric indicators at the micro-level: some results in the area of natural resources at the Spanish CSIC. *Research Evaluation*, 14 (2), 110-120

Caldarelli, G., A. Erzan, and A. Vespignani (eds.) (2004). Virtual round table on ten leading questions for network research. *The European Physical Journal B 38*, 143-145.

Dorogovtsev, S.N. and J.F.F. Mendes (2002). *Advances in Physics* 51, 1079-1187.

Katz, J.S. (1999). The self-similar science system. *Research Policy*, 28, 501-517.

Katz, J.S. (2000). Scale independent indicators and research assessment. *Science and Public Policy*, 27(1), 23-26.

Katz, J.S. (2005). Scale-independent bibiometric indicators. *Measurement*, 3(1), 24-28.

Leicht, E.A., G. Clarkson, K. Shedden, and M.E.J. Newman (2007). Large-scale structure of time evolving citation networks. *The European Physical Journal B*, 59, 75-83.

Merton, R.K. (1988). The Matthew Effect in Science, II: Cumulative advantage and the symbolism of intellectual property. *Isis* 79, 606-623.

Price, D. de Solla (1976). A general theory of bibliometric and other cumulative advantage processes. *Journal of the American Society for Information Science* 27 (5), 292-306.

van Raan, A.F.J. (2004). Measuring Science. Capita Selecta of Current Main Issues. In: H.F. Moed, W. Glänzel, and U. Schmoch (eds.). Handbook of Quantitative Science and Technology Research. Dordrecht: Kluwer Academic Publishers, p. 19-50.

van Raan A.F.J. (2006a). Statistical properties of Bibliometric indicators: Research group indicator distributions and correlations. *Journal of the American Society for Information Science and Technology* 57 (3), 408-430.

van Raan, A.F.J. (2006b). Performance-related differences of bibliometric statistical properties of research researchers: cumulative advantages and hierarchically layered networks. *Journal of the American Society for Information Science and Technology* 57 (14), 1919-1935.

van Raan, A.F.J. (2008a). Scaling rules in the science system: Influence of field-specific citation characteristics on the impact of research groups. *Journal of the American Society for Information Science and Technology* 59 (4) 565-576.

van Raan, A.F.J. (2008b). Self-citation as an impact-reinforcing mechanism in the science system. *Journal of the American Society for Information Science and Technology*, to be published.

Rinia, E.J. Th.N. van Leeuwen, H.G. van Vuren, and A.F.J. van Raan (1998). Comparative analysis of a set of bibliometric indicators and central peer review criteria. Evaluation of condensed matter physics in the Netherlands. *Research Policy* 27, 95-107.

Rinia, E.J., Th.N. van Leeuwen, H.G. van Vuren, and A.F.J. van Raan (2001). Influence of interdisciplinarity on peer-review and bibliometric evaluations. *Research Policy* 30, 357-361.

**Appendix**

*Table A1: Productivity classes (number of publications/researcher) by area*

| | Biology & Biomedicine | Materials Science | Natural Resources |
|---|---|---|---|
| P1 | 5-15 | 5-27 | 5-14 |
| P2 | 16-24 | 28-42 | 15-23 |
| P3 | 26-36 | 43-62 | 23-34 |
| P4 | > 36 | >62 | >34 |

*Table A2: Statistical data on the analysis of the size-dependence advantage of different bibliometric indicators: a) correlation between bibliometric indicators and number of publications; b) comparison of means by productivity classes.*

| Indicator | Area | *FCSm* class | Correlation with *P* | ANOVA by productivity classes | |
|---|---|---|---|---|---|
| | | | *a* | Sign. level | Sign. in post-hoc test |
| ***C*** | All | Top | +0.98 | 0.000 | All significant |
| | | Bottom | +1.32 | 0.000 | All significant |
| | Biol./Biomed. | Top | +1.01 | 0.000 | All significant |
| | | Bottom | +1.23 | 0.000 | All significant |
| | Material Sc. | Top | +1.28 | 0.000 | All significant |
| | | Bottom | +1.48 | 0.000 | All significant |
| | Nat. Resources | Top | +1.28 | 0.000 | All signficant |
| | | Bottom | +1.38 | 0.000 | All signficant |
| ***CPP*** | All | Top | -0.02 | N.S | - |
| | | Bottom | +0.32 | 0.000 | P1 vs P2, P3,P4 |
| | Biol./Biomed. | Top | +0.01 | N.S. | - |
| | | Bottom | +0.23 | N.S. | - |
| | Material Sc. | Top | +0.28 | 0.05 | P4 vs P1, P2 |
| | | Bottom | +0.48 | 0.01 | P1 vs P2, P4 |
| | Nat. Resources | Top | +0.28 | 0.01 | P1 vs P4 |
| | | Bottom | +0.38 | 0.01 | P1 vs P2, P3, P4 |
| ***FCSm*** | All | Top | -0.05 | 0.000 | P1 vs P2, P3, P4 |
| | | Bottom | +0.02 | N.S. | - |
| | Biol./Biomed. | Top | -0.05 | 0.01 | P1 vs P2, P4 |
| | | Bottom | +0.09 | 0.05 | P1 vs P4 |
| | Material Sc. | Top | -0.01 | N.S. | - |
| | | Bottom | +0.16 | 0.01 | P4 vs P1, P3 |
| | Nat. Resources | Top | +0.01 | N.S. | - |
| | | Bottom | +0.06 | N.S. | - |
| ***JCSm*** | All | Top | -0.17 | 0.000 | P1 vs P2, P3, P4; P2 vs P4 |
| | | Bottom | +0.18 | 0.001 | P1 vs P2, P3, P4 |
| | Biol./Biomed. | Top | -0.06 | N.S. | - |
| | | Bottom | +0.04 | N.S. | - |
| | Material Sc. | Top | -0.02 | N.S. | - |
| | | Bottom | +0.31 | 0.05 | P1 vs P2, P4 |
| | Nat. Resources | Top | +0.08 | N.S. | - |
| | | Bottom | + 0.28 | 0.05 | P1 vs P4 |

***C***= Number of citations; ***CPP***= Citation per publication; ***FCSm***= Field Citation Score (mean); ***JCSm***= Journal Citation Score (mean). 'Top researchers' refers to top-25% in the ***FCSm*** distribution. 'Bottom researchers' refers to bottom-25% in the ***FCSm*** distribution.

*Table A3:**CPP** by the Crown Indicator (**CPP/FCSm**) classification.*

| Scientific Field | Crown Indicator | Productivity Class | N | Mean *CPP* | Sign. |
|---|---|---|---|---|---|
| Biology & Biomedicine | ***CPP/FCSm***>1 | P1 | 46 | 3.33 | 0.001 |
| | | P2 | 49 | 3.15 | |
| | | P3 | 44 | 2.96 | |
| | | P4 | 41 | 2.93 | |
| | | Total | 180 | 3.1 | |
| | ***CPP/FCSm***<1 | P1 | 48 | 1.76 | 0.001 |
| | | P2 | 50 | 1.97 | |
| | | P3 | 43 | 2.06 | |
| | | P4 | 50 | 2.16 | |
| | | Total | 191 | 1.99 | |
| Materials Science | ***CPP/FCSm***>1 | P1 | 27 | 1.81 | N.S. |
| | | P2 | 18 | 1.89 | |
| | | P3 | 30 | 2.06 | |
| | | P4 | 39 | 1.95 | |
| | | Total | 114 | 1.93 | |
| | ***CPP/FCSm***<1 | P1 | 51 | 0.7 | 0.000 |
| | | P2 | 61 | 0.95 | |
| | | P3 | 45 | 0.91 | |
| | | P4 | 31 | 1.08 | |
| | | Total | 188 | 0.89 | |
| Natural Resources | ***CPP/FCSm***>1 | P1 | 20 | 2.02 | N.S. |
| | | P2 | 30 | 2.04 | |
| | | P3 | 26 | 2.1 | |
| | | P4 | 33 | 2.03 | |
| | | Total | 109 | 2.05 | |
| | ***CPP/FCSm***<1 | P1 | 62 | 0.85 | 0.000 |
| | | P2 | 52 | 1.26 | |
| | | P3 | 39 | 1.31 | |
| | | P4 | 42 | 1.42 | |
| | | Total | 195 | 1.18 | |

For researchers with a ***CPP/FCSm*** lower than 1, ***CPP*** tends to increase with production (***P***), so 'expanding in size' means higher ***CPP*** scores (differences statistically significant in all scientific fields). For researchers with ***CPP/FCSm*** higher than 1, no statistical differences in ***CPP*** by productivity class were found, except in Biology & Biomedicine, where ***CPP*** tends to decrease for the most productive researchers.